\documentclass[12pt,preprint]{aastex61}
\usepackage[utf8]{inputenc}
\usepackage{xcolor}
\usepackage{graphicx}
\usepackage{natbib}
\usepackage{amsmath}
\RequirePackage{lineno}
\renewcommand{\thefootnote}{\fnsymbol{footnote}}

\def\apjs{{ApJS}}
\def\apjl{{ApJL}}

\def\aap{{A\&A}}

\newcommand{\beginsupplement}{%
        \setcounter{table}{0}
        \renewcommand{\thetable}{S\arabic{table}}%
        \setcounter{figure}{0}
        \renewcommand{\thefigure}{S\arabic{figure}}%
     }

\begin{document}

\title{Stellar Chemical Clues as to The Rarity of Exoplanetary Tectonics}

\author{Cayman T. Unterborn}
\altaffiliation{SESE Exploration Fellow}
\affiliation{School of Earth and Space Exploration, Arizona State University, Tempe, AZ 85287, USA}
\email{cayman.unterborn@asu.edu}

\author{Scott D. Hull}
\affiliation{School of Earth Sciences, The Ohio State University, Columbus, OH, 43210}

\author{Lars Stixrude}
\affiliation{Department of Earth Sciences, University College London, London, UK}

\author{Johanna K. Teske}
\altaffiliation{Carnegie Origins Fellow, jointly appointed between Carnegie DTM \& Observatories}
\affiliation{Carnegie Observatories, 813 Santa Barbara Street, Pasadena, CA 92201}

\author{Jennifer A. Johnson}
\affiliation{Department of Astronomy, The Ohio State University, Columbus, OH, 43210}

\author{Wendy R. Panero}
\affiliation{School of Earth Sciences, The Ohio State University, Columbus, OH, 43210}

\date{June 2017}

%%%%%%%%%%%%%%%%%%%%%%%%%%%%%%%%%%%%%%%%%%%%%%%%%%%%%%%%%%%%%

\begin{abstract}

Earth's tectonic processes regulate the formation of continental crust, control its unique deep water and carbon cycles, and are vital to its surface habitability. A major driver of steady-state plate tectonics on Earth is the sinking of the cold subducting plate into the underlying mantle. This sinking is the result of the combined effects of the thermal contraction of the lithosphere and of metamorphic transitions within the basaltic oceanic crust and lithospheric mantle. The latter of these effects is dependent on the bulk composition of the planet, e.g., the major, terrestrial planet-building elements Mg, Si, Fe, Ca, Al, and Na, which vary in abundance across the Galaxy. We present thermodynamic phase-equilibria calculations of planetary differentiation to calculate both melt composition and mantle mineralogy, and show that a planet's refractory and moderately-volatile elemental abundances control a terrestrial planet's likelihood to produce mantle-derived, melt-extracted crusts that sink. Those planets forming with a higher concentration of Si and Na abundances are less likely to undergo sustained tectonics compared to the Earth. We find only 1/3 of the range of stellar compositions observed in the Galaxy is likely to host planets able to sustain density-driven tectonics compared to the Sun/Earth. Systems outside of this compositional range are less likely to produce planets able to tectonically regulate their climate and may be inhospitable to life as we know it.
\end{abstract}
%\keywords{}

%%%%%%%%%%%%%%%%%%%%%%%%%%%%%%%%%%%%%%%%%%%%%%%%%%%%%%%%%%%%%

\section{Introduction}
\modulolinenumbers[5]
%\linenumbers
The Earth is unique in our Solar System. It is the only planet with plate tectonics and liquid water on the surface. It is not known, however, the extent to which the Earth is unique among all terrestrial planets beyond our Solar System. The \textit{Kepler} mission's discoveries establish that Earth-sized planets are common in the Galaxy, with as many as 11\% of Sun-like stars hosting planets 1-2 times the radius of the Earth and receiving comparable solar flux \citep{Marc14,Fult17}. Together with other discovery campaigns, we know now of many exoplanets with masses and radii consistent with being terrestrial, rock/metal-dominated planets, rather than gas-dominated. The degree to which these planets can maintain surface oceans, plate tectonics or even be considered ``Earth-like'' is not known and is a complex function of the planet's composition, formation, and dynamical state \citep[e.g. ][]{Fole15,Fole16}. We assert that for a planet to be ``Earth-like'' and habitable, it must be habitable in the same manner as the Earth. At a minimum this means the planet must sustain surface liquid water for millions to billions of years. . Because stable liquid water exists in a relatively narrow range in temperatures and pressures, the planet must have a process to regulate atmospheric temperatures. On the Earth, moderate temperatures are maintained by the incoming solar radiation combined with moderate greenhouse warming from CO$_2$, H$_2$O, and CH$_4$. Therefore, supply and regulation of these gases is key to Earth's, and thus ``Earth-like'' planet's climate. This definition is in contrast to the typical one for ``Earth-like," in which a planet is defined simply as one with a bulk density characteristic of being roughly that of a mixture of metal $\pm$ rock.

The Earth's atmospheric regulatory processes and aqueous chemistry arise from tectonics: the recycling of material between a planet's surface and mantle. For Earth, the recycling process manifests as special case of tectonics, plate tectonics, in which oceanic crust continuously subducts into the interior and convective upwellings returning some material to the surface. This Earth-scale transport of material produces the buoyant continental crust and releases and sequesters CO$_2$ through weathering of silicate rocks and arc volcanism at subduction zones \citep{Velb93,Brad91,Slee01,Fole15,Fole16}. In contrast a terrestrial planet without tectonic processes, such as one with a rigid lid like Mars, or undergoing episodic overturn like Venus, does not have a steady state cycling of material between the surface and interior. Even if Venus had a lower surface temperature, it is unable to regulate atmospheric CO$_{2}$, which may be rapidly released in pulses during overturn or a consequence of its solidification from a magma ocean \citep{Hama13}. The lack of standing continents and the formation of surface carbonic acid on Venus does not allow CO$_{2}$ to be efficiently buried, and it instead accumulates in the atmosphere, creating a runaway greenhouse.

The planetary controls that lead to mobile-lid regimes, including plate tectonics, and static-lid regimes are a matter of debate even for the well-determined compositional, thermal, and structural parameters found in the Solar System \citep[e.g.][]{More98}. Models of large, terrestrial planets (so-called Super-Earths) have concluded that plate tectonics could be inevitable due to interior-to-surface heat transfer, surface gravity, and fault strength \citep{Vale07a,vanH11}, while other models, focused on the fault strength integral to subduction initiation, have come to the opposite conclusion that plate tectonics are unlikely  \citep[e.g.][]{ONei07}. More general models find the tectonic state of a super-Earth is a function of planet size, incident solar radiation, and atmospheric composition \citep[e.g.][]{Fole12}. Each of these models, however, simply scale the Earth in composition and structure and sought only to understand how changes in the physical parameters of a planet affected tectonics, rather than address the much more complicated (and data-lacking) question of how planet chemical diversity affects tectonics.

An alternate approach is to assess \textit{probabilistically}, rather than \textit{definitively}, the relative likelihood of tectonics on exoplanets in individual systems by examining the effect of planetary chemistry on plate tectonics \citep{Unte14,Unte15,Stam16}. In this study, we quantify the effect of planet composition on a vital aspect of sustaining plate tectonics on terrestrial planets over billion year timescales: the sinking forces of the exocrust into a planet's mantle. We therefore address a minimum criterion for plate tectonics: whether or not the surface crusts sink into the mantle due to buoyancy forces arising from thermal and chemical differences between the surface crust and the interior mantle. In the most general sense, tectonic processes of both the Earth and Venus are driven by buoyancy forces. The magnitude of this force is proportional to the integrated density difference between the sinking surface material and the surrounding mantle. These density contrasts are due to the composition and thermal state of the surface material compared to the mantle. Those planetary compositions in which crustal material is buoyant even when forced downward through crustal thickening or contraction have no mechanism for cycling crustal material into the interior via subduction. For those planets that have a buoyancy force less than the Earth, though, there is a lower probability of plate tectonics compared to Earth due to the reduced negative buoyancy (that is downward force) available to drive plate tectonics. Those systems are \textit{less likely} to produce planets with lithospheric material of a composition able to sink via buoyancy forces to $\sim$100 km and thereby produce arc volcanism, will not be able to maintain long-lived temperate, atmospheres, nor will they exhibit top-down, buoyancy driven, steady-state crustal recycling as part of tectonic processes. 

The Earth's subducting lithosphere is composed of two parts: a 5-10 km thick basaltic layer lying on top of a 50-80 km thick, cold, and rigid layer of lithospheric mantle \citep[e.g.][]{Fisc10}, which formed as a result of the cooling of the surface of Earth followed by induration from a magma ocean for 50-200 million years. The basaltic layer is formed through eruption at mid-ocean ridges as the result of decompression melting of the passive upwelling of the convecting mantle. The composition of oceanic lithosphere is therefore controlled by the temperature and composition of the mantle. 

Once the subducting basaltic crust reaches a depth of 35-50 km below the surface, pressure-induced mineral metamorphism transforms the basaltic rock into rock denser than the surrounding mantle ($\Delta\rho\sim+100$ kg m$^{-3}$) through the formation of garnet as it replaces orthopyroxene with minor spinel \citep[][ Supplemental Figure \ref{fig:phases}]{Fros08}. This basalt-eclogite transition marks the first of two major metamorphic processes in the basaltic layer responsible for the negative buoyancy that causes plates to sink and continue the cycling from surface to mantle and back. The second phase transition occurs at 300 km below the surface, with excess silica (SiO$_2$) undergoing the coesite-stishovite transition, providing an additional density increase, further promoting sinking. The subducting plate is also thermally contracted relative to the surrounding mantle. The resulting density contrast produces an additional downward buoyancy force, controlled by the temperature difference between subducting plate and surrounding mantle. 

Pressure- and temperature-induced phase transitions also occur within the mantle surrounding the subducting plate. However, because of the different temperatures of the mantle relative to the plate, these transitions happen at different depths for each. At a depth near 410 km within the Earth, a phase transition occurs in mantle olivine [(Mg$^{2+}$,Fe$^{2+}$)SiO$_4$], transforming it to wadsleyite. This phase transition is what delineates Earth's upper mantle from the transition zone (Supplementary Figure \ref{fig:phases}). The depth of transition from olivine to wadsleyite is shallower at lower temperatures such that olivine in the relatively colder subducting plate will transform into the denser wadsleyite at depths shallower than the transition zone. This introduces a wedge of more dense material in the sinking plate above the 410 km transition, further promoting the sinking of subducting plates. 

These phase changes, and the relative depths of their occurrence, result in a chemical buoyancy force, $F_{c}$, on the plate. The integrated density contrast between the plate and mantle due to thermal differences, including depth to pressure-induced transitions, results in a thermal buoyancy force, $F_{t}$. High-temperature atmospheres limit the cooling of the plate on the surface \citep{Fole16}, which will reduce the magnitude of $F_{t}$ and lower the likelihood of plate tectonics on these planets. To date, there are no definitive observations of terrestrial exoplanet atmospheres and thus inference of mobile versus stagnant-lid states through atmospheric observation is currently beyond the reach of exoplanetary science. 

Together, $F_{c}+F_{t}$ for Earth's subducting plates contribute a net buoyancy of  $\sim-(2-3)* 10^{13}$ N per meter length of subducted plate downward \citep[recalculated from ][]{Kird14}. While the processes responsible for the initiation of subduction are not well-understood on Earth, plates having sufficient negative buoyancy at the depth of the basalt-eclogite transition is a necessary precondition for incipient subduction and prevention of a stagnant lid \citep{vanH02}. These upper mantle forces, together with the shear traction and induced flow of a slab inciting downward mantle motion, constitute one of two major driving forces of plate tectonics \citep{Conr02}.

An exoplanet's composition and mineralogy are not directly observable and are instead inferred from mass-radius relations \citep[e.g.][]{Seag07,Weis14,Wolf16,Unte16}. Given only a planet's mass and radius, there is considerable degeneracy in planet composition and structure. As noted by \citet{Dorn15}, this degeneracy is reduced when the stellar composition of terrestrial planet-building elements (Mg, Si, Fe) is adopted as a proxy for planetary composition. This is well grounded for the Sun-Earth system due to the refractory nature of these elements \citep[][, Supplementary Table \ref{tab:comparison}]{Unte16} and has yet to be tested for Venus as there are no definitive measurements of its interior composition or structure. Relative to solar \citep{Aspl05}, the abundances of the major terrestrial planet-building elements Mg, Si, and Fe vary between 10 and 400\% of solar in large stellar surveys \citep[e.g.][]{Adib12,Bens14,Hink14,Brew16}. Variations in stellar (Mg+2Si)/Fe affect an exoplanet's core mass fraction and the ratio Mg/Si affects the dominant minerals of the silicate mantle. Decreasing Mg/Si will lead to a shift in the dominance of the upper-mantle mineral Mg$_2$SiO$_4$ (olivine), to MgSiO$_3$ (pyroxene), to SiO$_2$ \citep[quartz;][]{Unte16,Unte17}. While oxidation state of a planet will affect each of these simple assumptions somewhat, the extent of these oxidation-reduction reactions is a complex function of disk chemistry and the planet's initial  thermal profile and subsequent evolution, each of which are poorly understood areas in exoplanetary science. To first order, though, the assumption that stellar refractory composition roughly mirrors planetary refractory composition provides testable predictions of planetary mass for those planets orbiting stars of non-Solar composition \citep{Unte17}. 

We therefore adopt the stellar compositional diversity observed in the Galaxy as a proxy for the compositional diversity of terrestrial planets. From this starting point, we calculate the composition, mineralogy of the melt-extracted crust and residual mantle as relative contrast in buoyancy between each. We thus quantify how the composition of a rocky planet affects the likelihood of sustaining plate tectonics over billion-year timescales. The melt-extracted crust is created when solid-state convection develops in the bulk silicate planet (BSP) after the differentiation of a core and the adiabatic rise of rock towards the surface leads to partial melting of the mantle, producing a melt-extracted crust of different composition from the initial bulk silicate composition. A key feature of continual plate tectonics over geologic timescales is this melt-extracted crust subducting along with underlying lithospheric mantle into the interior. The relative magnitude of buoyancy contrast between the melt-extracted crust and residual mantle, therefore, will lead to a greater or lesser \textit{likelihood} of plate tectonics in comparison to the Earth. 

\section{Results}
\begin{figure*}
    \centering
    \includegraphics[width=\linewidth]{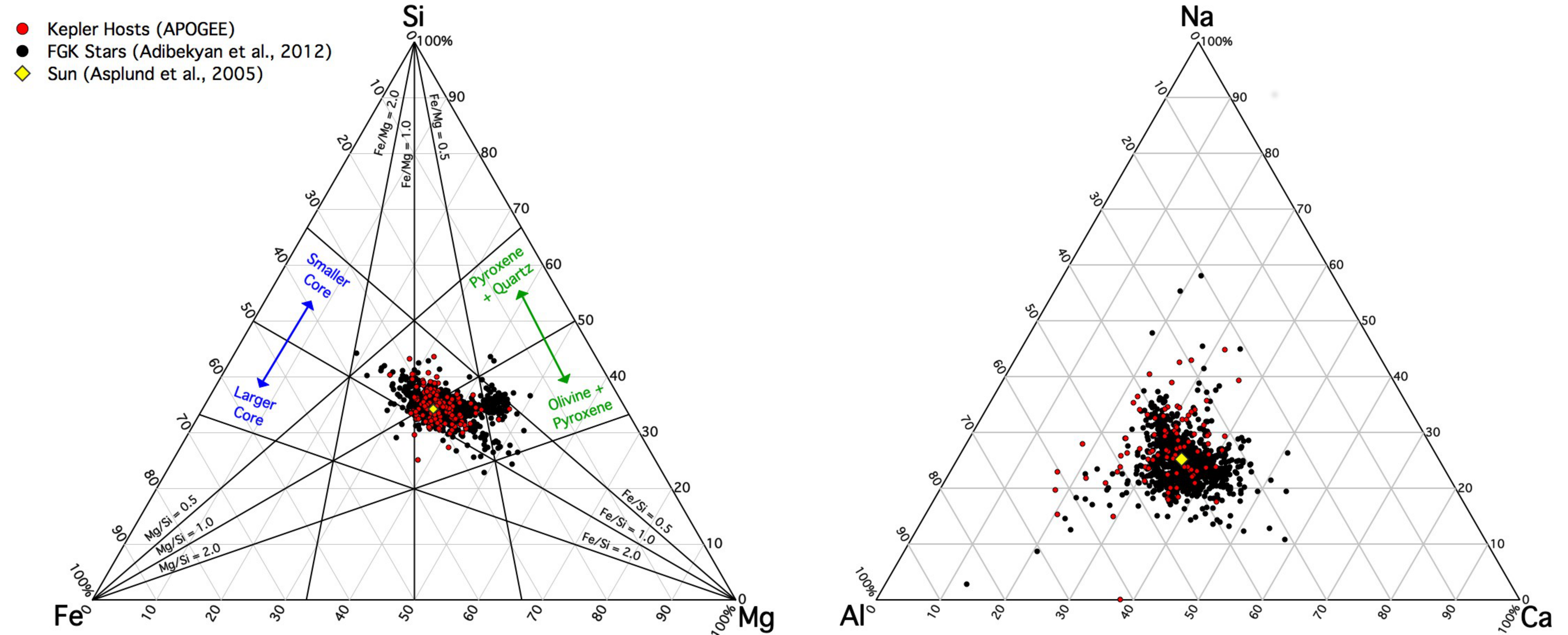}
    \caption{Ternary diagrams of stellar Mg, Fe and Si \& Al, Ca and Na molar abundances for our adopted sample of FGK stars \citep[black,][]{Adib12} (black, Adibekyan et al., 2012;) and \textit{Kepler} host stars found in the APOGEE survey DR13 \citep[red,][]{Alba16} and the yellow diamond is Solar \citep{Aspl05}. The \citet{Adib12} database is based on spectra of resolution R $\sim$ 110,000 and signal-to-noise ratio ranging from $\sim$20 to $\sim$2000. Fifty-five percent of the spectra have a signal-to-noise greater than 200, with only 17\% lower than 100. The average observational uncertainty is ~25\% in relative abundances, well inside the range observed abundance variation in this sample. The APOGEE data set is based on spectra of resolution R $\sim$22,000 and signal-to-noise ranging from $\sim$30 to $\sim$900. Seventy-five percent of the spectra have signal-to-noise greater than 100, with only 3\% percent lower than 50. While our sample is scaled from solar abundances, changes in these abundances will affect the abundances of each star uniformly and thus the observed elemental ratios and relative buoyancy forces would not change across our sample.}
    \label{fig:ternaries}
\end{figure*}

\begin{figure}
    \centering
    \includegraphics[width=.6\linewidth]{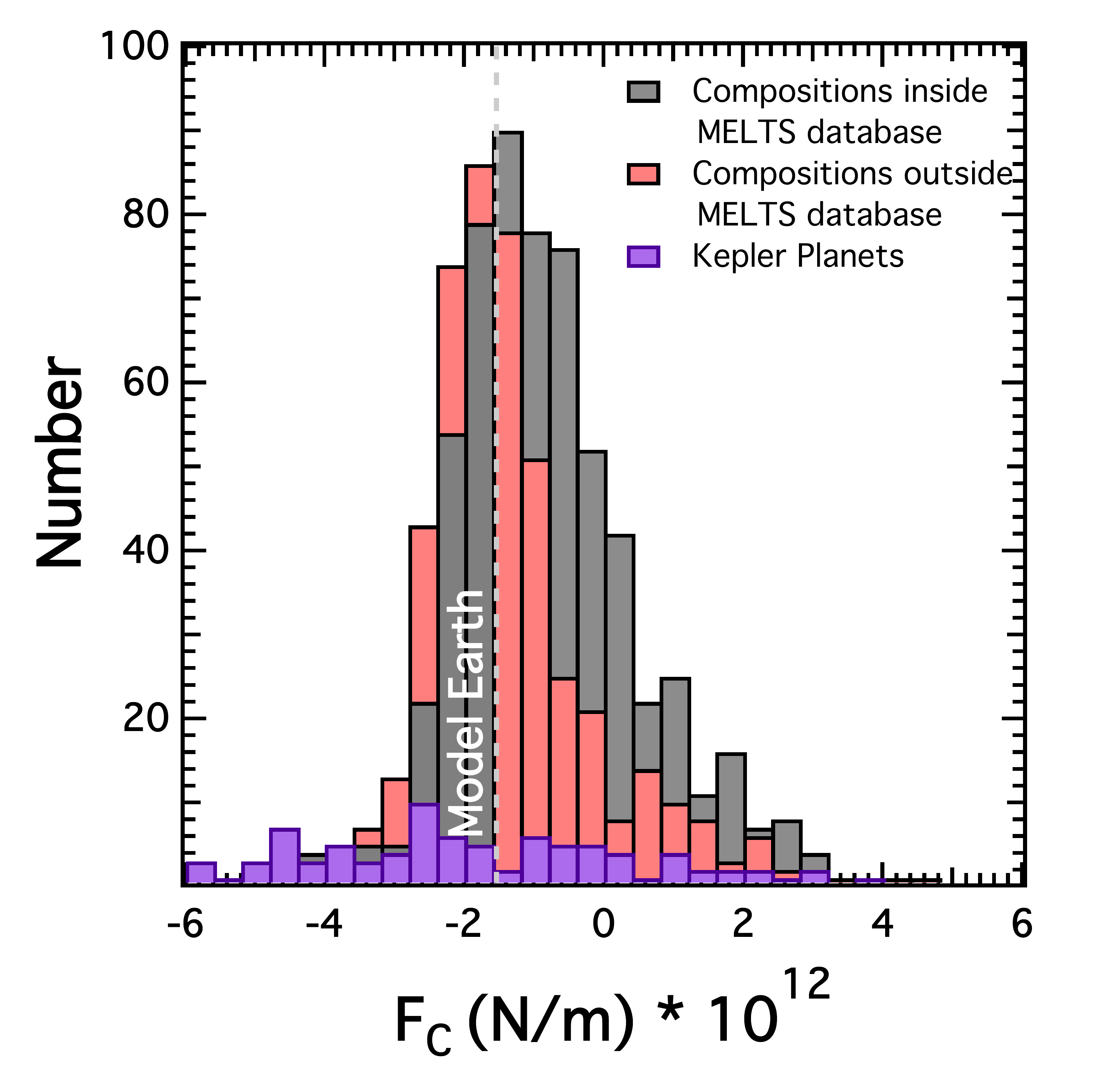}
    \caption{Histograms of buoyancy forces calculated using our model for our sample of 609 \citet{Adib12} stars (gray) and 89 \textit{\textit{Kepler}} planet hosts (purple) with compositions inside the MELTS database \citep[][]{Ghio02}. Those modeled from stellar composition are shown in red. The buoyancy force and bulk composition of model Earth is shown as a light gray dashed line.}
    \label{fig:histogram}
\end{figure}

We calculate the melt composition and stable mineralogy of the bulk silicate planet and exocrust, benchmarking our model to the Earth as derived from solar (See Supplementary Materials; Supplementary Figure \ref{fig:phases}) for hypothetical planetary compositions derived from two samples of stellar compositions. The first includes 1063 FGK stars from \citet{Adib12}, which represent thin and thick-disk stars with a metallicity range from -0.8 $<$ [Fe/H]\footnote[2]{[Fe/H]= $log \left(N_{\rm{Fe}}/N_{\rm{H}}\right)_{\rm{star}}-log \left(N_{\rm{Fe}}/N_{\rm{H}}\right)_{\rm{Sun}}$, where $N_{\rm{Fe}}$ and $N_{\rm{H}}$ are the number of atoms of Fe and H respectively} $<$ 0.6, over 100 of which are known to host (mostly gas-dominated) planets. Of this first sample, we thermodynamically model the composition and buoyancy of a subset of 609 stellar compositions that are within the internally consistent MELTS database \citep{Ghio02}, 57 of which are known to host planets. We find a chemical buoyancy ($F_c$) of these systems ranges from $-7.5*10^{12}$ N m$^{-1}$ (sinks) to $4.0*10^{12}$ N m$^{-1}$ (floats) while our model Earth produces buoyancy force of $\sim-2.0*10^{12}$ N m$^{-1}$ (Figure \ref{fig:histogram}).

The second sample of stellar compositions comes from the Apache Point Observatory Galactic Evolution Experiment \citep[APOGEE,][]{Wils10,Maje15}, which is part of the Sloan Digital Sky Survey IV \citep[SDSS-IV;][]{Blan17}. From the public APOGEE Data Release 13 \citep[DR13,][]{Alba16}, we selected the stars known to host small planets (R$_{\rm {p}}$ $\leq$1.6 R$_{\oplus}$) from the \textit{Kepler} transiting planet survey, totaling 123 stars with a metallicity range from -0.54 $<$ [Fe/H] $<$ 0.36. Of these, we thermodynamically model the composition and buoyancy of a subset of 89 stellar compositions that are within the internally consistent MELTS database \citep{Ghio02}. We find a chemical buoyancy of these systems ranges from $-10*10^{12}$ N m$^{-1}$ (sinks) to $3.6*10^{12}$ N m$^{-1}$ (floats).

Of terrestrial planets with compositions represented by the variation in both stellar datasets, 19\% would produce exocrusts entirely buoyant than their BSP throughout the entire upper mantle ($F_{c} > 0$), and therefore unable to subduct (Figure \ref{fig:histogram}). Furthermore, 61\% of the Adibekyan sample and 41\% of the \textit{Kepler} sample produces plates more chemically buoyant than that of our model Earth ($F_{c} > F_{c-Earth}$; Figure \ref{fig:histogram}) and therefore \textit{less likely} to host plate tectonics than the Earth. 

\begin{figure}
    \centering
    \includegraphics[width=.6\linewidth]{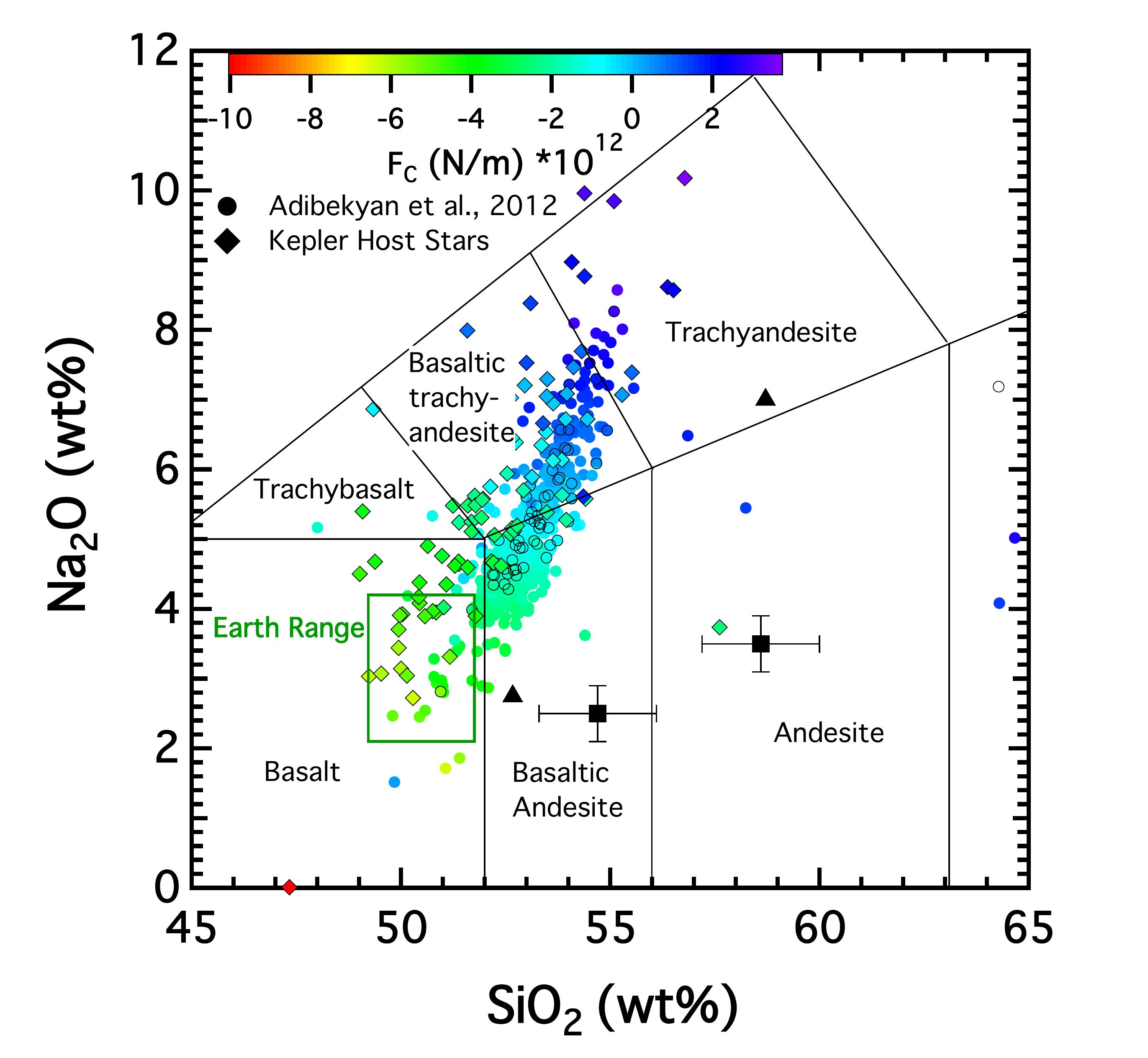}
    \caption{Alkali abundance as a function of silica of the exocrust compositions for our sample of \textit{Kepler} planet-host (diamonds) and \citet{Adib12} stars (circles). Color is a function of the chemical buoyancy as calculated from a thermally equilibrated, 5\% melt layer. Points outlined are those with known planets (exoplanets.org). The green box represents the Earth Range of average Earth basaltic compositions of \citet{Gale13}. The composition of Mercury's Northern Volcanic Plains and Intercrater Plains and Heavily Cratered Terrains \citep[diamonds;][]{Namu16} and Mars' type 1 and type 2 crusts \citep[squares, including K$_2$O;][]{McSw03} are included. Venus' alkali composition has yet to be determined \citep{Trei13}. Each crustal composition of Mercury and Mars fall within the positively buoyant andesitic field, despite 3 of the 4 measurements containing Na$_2$O abundances within the Earth-like range.}
    \label{fig:rocktype}
\end{figure}

This variation in buoyancy forces is mainly caused by the enrichment of Si and alkali elements (Na, K) within the melt-extracted crust. Partial melting of typical mantle rocks leads to enrichment of silica and alkalis in the melt compared to the parent rock. We find those planetary compositions with lower alkali and silica abundance, such as those crustal materials that form on Earth with basaltic composition, are most likely to sink, while those with greater alkaline and silicic compositions are less likely (Figure \ref{fig:rocktype}; Supplementary Figure \ref{fig:nonEarth}). This is consistent with Earth's tectonics, in which low-density and alkali-rich, andesitic continental crust does not subduct \citep{Cloo93}. 

The magnitude of chemical buoyancy is, therefore, closely tied to the abundance of sodium and potassium in the BSP due to their incompatibility in melting processes (Figure \ref{fig:histogram}). Because Na is roughly 15 times more abundant than K for the Sun \citep{Lodd03}, we focus on Na as a controlling alkali in the buoyancy calculations. Because all alkalis are moderately volatile in the planetary condensation process \citep{Lodd03}, their abundances are not likely to be mirrored in any orbiting terrestrial planets due to their fractionation relative to the refractory elements (e.g. Mg, Si, Fe) in the planetary formation process. For this model, we assume an incomplete accretion of Na due to volatile loss during planet formation in proportions similar to Sun-Earth fractionation for each star in our sample \citep[$\delta_{ \rm{Na-Sun/Earth}} = \left (\rm{Na/Si}\right)_{\rm{Earth}}/\left(\rm{Na/Si}\right)_{Sun}$ = 0.26 by mole; ][]{McD03,Aspl05}. Chemical buoyancy is then a function of the total Na in the star as well as the degree of Na volitalization during accretion, $\delta_{\rm {Na}}$:
\begin{equation}
\label{eq:fit}
F_{c} \approx -1.15+16.6\left(\rm{[Na/Si]}+log10\left(\delta_{\rm{Na}}/\delta_{\rm{Na-Sun/Earth}}\right)\right)  * 10^{12} \rm{N m^{-1}}
\end{equation}

\begin{figure}
    \centering
    \includegraphics[width=\linewidth]{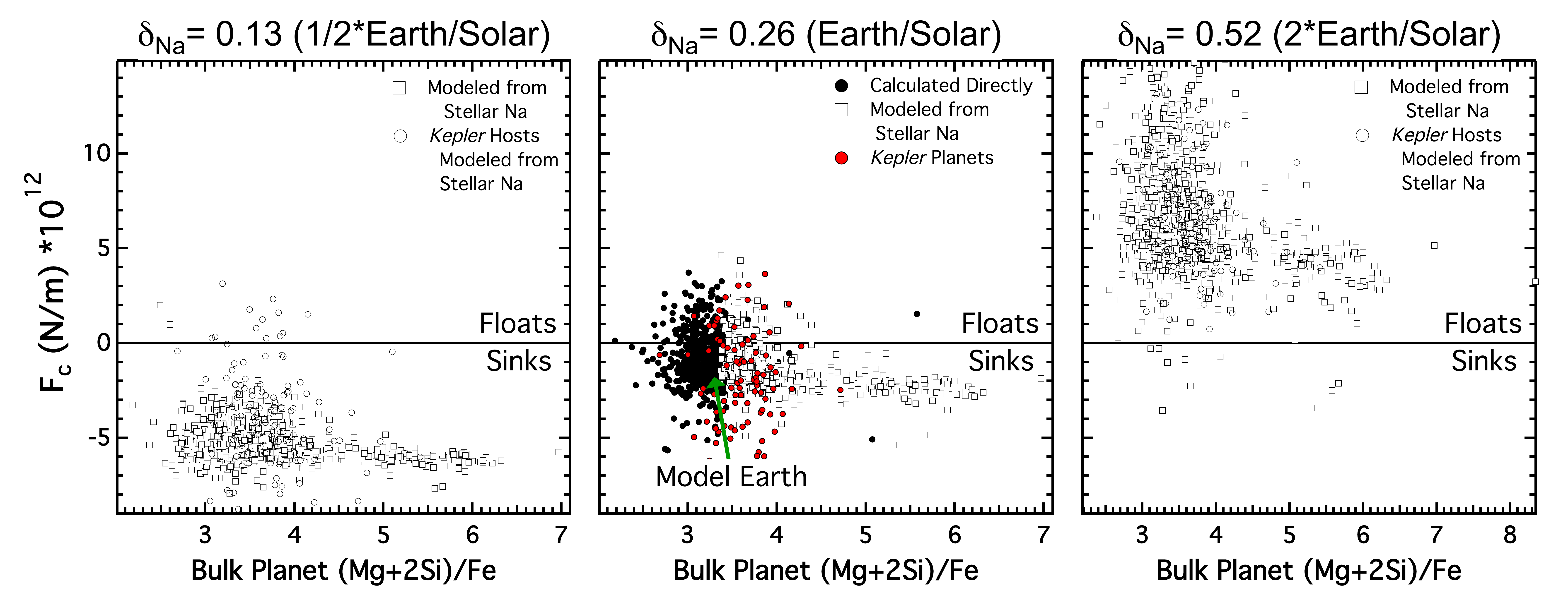}
    \caption{Chemical buoyancy force ($F_{c}$) as a function of bulk planetary Mg and Si normalized to Fe (by mole) for values for three values of $\delta_{\rm{Na}}$. Filled circles are those calculations presented here in Figures \ref{fig:histogram} and \ref{fig:rocktype} for the \citet{Adib12} dataset (black) and \textit{Kepler} hosts (red). Open symbols are those buoyancy forces calculated from the relationship between stellar [Na/Si] and $F_{c}$ (Equation \ref{eq:fit}, Supplementary Figure \ref{fig:fit}) for the \citet{Adib12} stars (squares) and \textit{Kepler} hosts (circles). The 1063 stars in the stellar survey of \citet{Adib12} show $\sim$2.5 orders of magnitude difference in total Na, halving or doubling $\delta _{\rm {Na}}$ from the Solar value represents the entire sample either producing none and all plates being negatively chemically buoyant. The index (Mg+2Si)/Fe is both a primary control on the compositional extent of the thermodynamic database \citep{Ghio02} as well as defining the approximate proportions of core and mantle of an associated terrestrial planet \citep{Unte16}.  
}
    \label{fig:Sodium}
\end{figure}
when [Na/Si] is scaled from \citet{Aspl05} (Supplementary Figure \ref{fig:fit}), 78\% of the variation in $F_c$ can be explained by variation in [Na/Si]. For a given stellar [Na/Si] value, 95\% of $F_c$ values fall within $\pm10^{12}$ N m$^{-1}$, which is a more complex function of the refractory element composition of the planet and must be derived through the methods outlined here. This correlation between $F_c$ and Na abundance allows us to estimate the chemical buoyancy of the 454 stars in the \citet{Adib12} dataset whose mantle compositions were not located within the MELTS database and were generally those stars with bulk Fe/Si $<$ 0.95 \citep[Bulk Earth Fe/Si = 1;][]{McD03}. For those stars in the Adibekyan dataset, we find 28\% are of compositions likely to produce melt-extracted crust that remains buoyant relative to their mantles ($F_c > 0$), with 69\% less chemically buoyant than the Earth (Figure \ref{fig:histogram}). The Kepler dataset shows a similar trend, with 22\% of stars likely to produce crusts with positive chemical buoyancy and 60\% with crustal buoyancy forces greater than the Earth. Each of these systems are therefore \textit{less likely} than the Earth or \textit{completely unlikely} to maintain long-term, steady-state tectonics.

When $\delta_{\rm{Na}}$ is twice $\delta_{\rm{Na-Sun/Earth}}$ ($\delta_{\rm{Na}}$ = 0.52), only 1\% of stars within the Adibekyan sample and 4\% of the Kepler sample produce any plates with a negative chemical buoyancy force ($F_c < 0$), whereas all but fourteen stellar compositions in our sample of 1186 stars produce negatively buoyant plates when $\delta_{\rm{Na}}$ is half of $\delta_{\rm{Na-Sun/Earth}}$ ($\delta_{\rm{Na}}$ = 0.13; Figure \ref{fig:Sodium}). The likelihood of producing negative chemical buoyancy in melt-extracted surface plates is therefore very sensitive to $\delta _{\rm {Na}}$ and weakly dependent upon bulk mantle composition. Stars with greater Na abundance than Solar must retain less Na when forming terrestrial planets to create negatively buoyant plates at depth. The same is true for those stars of Solar Na composition with greater $\delta _{\rm {Na}}$ compared to the Earth and Sun ($\delta _{\rm {Na}} >$ 0.26). These high Na planetary systems may initiate subduction, but due to their higher positive chemical buoyancy force, they are unlikely to continue to do so \citep{Cloo93}. As mantle temperatures increase in this stagnant-lid planet, any volatiles contained within the partially subducted plate will degass into the atmosphere,  thus entirely negating any long-term, continuous tectonic regulation of climate.  

Our compositionally focused approach is a tool to examine the influence of a host star's refractory and moderately volatile composition on a key aspect of planetary dynamics: top-down, density-driven plate tectonics. This model attempts to capture only the most general details of terrestrial planetary formation, differentiation, and evolutionary process with the sinking of a surface plate through a planet's mantle being a consequence of the difference in density between the mantle and crust. This model treat the complex processes of accretion, formation of overlying continents, and prolonged geochemical processing as discrete processes at a fixed pressure, temperature, and oxygen fugacity. While not meant to be strictly prescriptive, these results demonstrate first-order trends and controlling factors in the geochemical consequences of variable compositions of terrestrial planets. 

This approach predicts a lack of plate tectonics on Venus and the potential for (transient) plate tectonics on Mars. For planets with significant thermally insulating atmospheres, more heat is retained at the surface, limiting the cooling of warming the plate \citep{Fole15,Fole16}. A surface temperature 450 K greater than Earth such as Venus will significantly reduce $F_{t}$, while $F_{c-Venus} \sim F_{c-Earth}$, such that the sum predicts Venus is less likely than Earth to undergo tectonics without having to consider the strength or thickness of surface rocks \citep{Fole12,Berc14}. In the case of Mars, the lack of water and rapid heat loss due to its small size limits the likelihood for long-term plate tectonics, although it may have experienced subduction in the past \citep{Slee94}. 

One of the most significant neglected variables is the abundance of water and carbon in the planet's interior, which are sensitive to nebula composition, formation processes and location in the nebula, and subsequent evolution. The volatile abundance in terrestrial planets is not well constrained in planetary formation models, both for our Solar System and exoplanetary systems. While planetary H$_2$O and CO$_2$ abundances may inform the scale of exoplanetary oceans and climate, the planet's relative abundance of moderately volatile elements such as Na is a first-order control on the likelihood of regulating these important aspects of exoplanet habitability via long-term plate tectonics. This approach, therefore, points to the importance of addressing the thermal and dynamic controls for both the highly and moderately volatile abundances of planets in formation models.

Of the small exoplanets discovered to date (R $\leq$ 1.6 Earth radii), few have both size and mass measurements, from which mean density may be calculated. Mean density is a non-unique function of the relative proportions of core, mantle, and gaseous envelope. Of those mass/radius measurements sufficiently dense to suggest the planets are terrestrial, planetary compositions and structures are indistinguishable from one another due to large observational uncertainties. We offer a complementary approach for those planets inferred to be terrestrial by providing a framework for quantifying an exoplanet's potential geochemical dynamics. Indeed, these results represent the first observational metric with which to gauge the probability of an exoplanetary system to maintain steady state surface-to-interior geochemical cycling and tectonic behavior with the potential to regulate atmospheric temperatures over long timescales. Furthermore, the connection between stellar Na abundance, degree of volatilization, and surface plate chemical buoyancy point to the importance of adopting a comparative tool to ascertain the likelihood of exoplanetary plate tectonics compared to the Earth. The exoplanetary field needs to refine models of not only the volatile condensation and accretion in planetary systems, but track the fractionation and dynamics of the refractory and moderately volatile elements within the disk as well. Furthermore, geochemical and geophysical experiments and models must be expanded to those compositions relevant to planets of non-Earth/Solar composition. The upcoming James Webb Space Telescope will provide opportunities for time-intensive observational studies to follow up terrestrial exoplanet atmospheric conditions, and our framework provides a strategy to identify those planetary systems most worthy of follow up observations. Bulk planet density is only one factor in determining whether a planet is ``Earth-like.'' It is only through this holistic approach to characterizing exoplanets, combining data and models from across scientific fields, that we will truly calibrate the likelihood of an extrasolar planet to be behaviorally Earth-like and habitable to life as we know it.

\acknowledgments
We thank the Cooperative Institute for Dynamic Earth Research (CIDER) for providing the space and intellectual environment for fostering this work as well as funds to support the BurnMan code. This project was supported by NSF-CAREER (EAR-09-55647) to Panero and the Shell Undergraduate Research Experience, OSU Undergraduate Research Scholarship, and Friends of Orton Hall grant to Hull. Funding for the Sloan Digital Sky Survey IV has been provided by the Alfred P. Sloan Foundation, the U.S. Department of Energy Office of Science, and the Participating Institutions. SDSS acknowledges support and resources from the Center for High-Performance Computing at the University of Utah. The SDSS web site is www.sdss.org. SDSS is managed by the Astrophysical Research Consortium for the Participating Institutions of the SDSS Collaboration including the Brazilian Participation Group, the Carnegie Institution for Science, Carnegie Mellon University, the Chilean Participation Group, the French Participation Group, Harvard-Smithsonian Center for Astrophysics, Instituto de Astrof\'{\i}sica de Canarias, The Johns Hopkins University, Kavli Institute for the Physics and Mathematics of the Universe (IPMU) / University of Tokyo, Lawrence Berkeley National Laboratory, Leibniz Institut f\"{u}r Astrophysik Potsdam (AIP), Max-Planck-Institut f\"{u}r Astronomie (MPIA Heidelberg), Max-Planck-Institut f\"{u}r Astrophysik (MPA Garching), Max-Planck-Institut f\"{u}r Extraterrestrische Physik (MPE), National Astronomical Observatories of China, New Mexico State University, New York University, University of Notre Dame, Observat\'{o}rio Nacional / MCTI, The Ohio State University, Pennsylvania State University, Shanghai Astronomical Observatory, United Kingdom Participation Group, Universidad Nacional Aut\'{o}noma de M\'{e}xico, University of Arizona, University of Colorado Boulder, University of Oxford, University of Portsmouth, University of Utah, University of Virginia, University of Washington, University of Wisconsin, Vanderbilt University, and Yale University.

\bibliography{main.bbl}

\newpage
\section{Supplementary Material}
\beginsupplement
\subsection{Stellar Composition Data}
We apply this Earth-benchmarked model to two samples of stars with abundances available for 9 elements (Mg, Fe, Si, Ca, Al, Ni, Ti, Cr and Na). First is the survey of F, G and K stars in the Galaxy of \citet{Adib12} and the other is a sample of 123 known planet-hosting stars whose abundances were available within the APOGEE Data Release 13 \citep[DR13][]{Alba16}. Where available, only those stars with C/O $<$ 0.8 were chosen. We use 1186 total stars in our model. These stars represent both thin and thick-disk stars with a metallicity range from -0.8 to 0.6 [Fe/H], of which 235 have detected planets. Of these stars, 698 have compositions that fall within the thermodynamic databases used in this study (Figure \ref{fig:histogram}). For those stars outside of the database, we estimate the crustal Na content and chemical buoyancy force through a linear fit of these values for only those 609 stars that were within the MELTS database and in the \citet{Adib12} dataset (Supplemental Figure \ref{fig:fit}).

We note that a separate publication is in progress (Teske et al. in prep) that will discuss the reliability of FGK dwarf star abundances produced by the APOGEE automated analysis pipeline \citep[ASPCAP; ][; Holtzmann et al. in prep, Jonsson et al, in prep]{Garc16}. Initial results (also discussed in Wilson et al. in prep) suggest that, for solar-type stars, the Fe, Mg, and Si abundances produced by ASPCAP are reliable at the $\sim$0.15 dex level. The ASPCAP Na abundances appear to be less reliable, upon comparison with a small sample of values derived from optical data. For consistency, we did not modify the DR13 APOGEE Na abundances to bring them into better agreement with the optical data. We note that Figure 1 shows that the \citet{Adib12} and the APOGEE samples have similar ranges in Na abundances.

\subsection{Methods Summary}
\subsubsection{Planet Composition Model}
There is extensive debate as to the exact nature of the Earth's composition, from one closer to a volatilized carbonaceous chondrite \citep{McD95,McD03}, to more similar to enstatite chondrites \citep{Javo10}, to mixtures of carbonaceous chondrites and ordinary chondrites \citep[e.g.][]{Fito12}. We instead adopt the simplest model of a composition that is identical to the Sun in refractory elements, with an oxygen abundance fixed by the fO$_{2}$ of the condensing medium. For key moderately volatile elements such as Na, we assume an ad-hoc condensation efficiency of 26\% to reproduce the Earth's bulk Na composition \citep{Aspl05,McD03}.

\begin{figure}
    \centering
    \includegraphics[width=\linewidth]{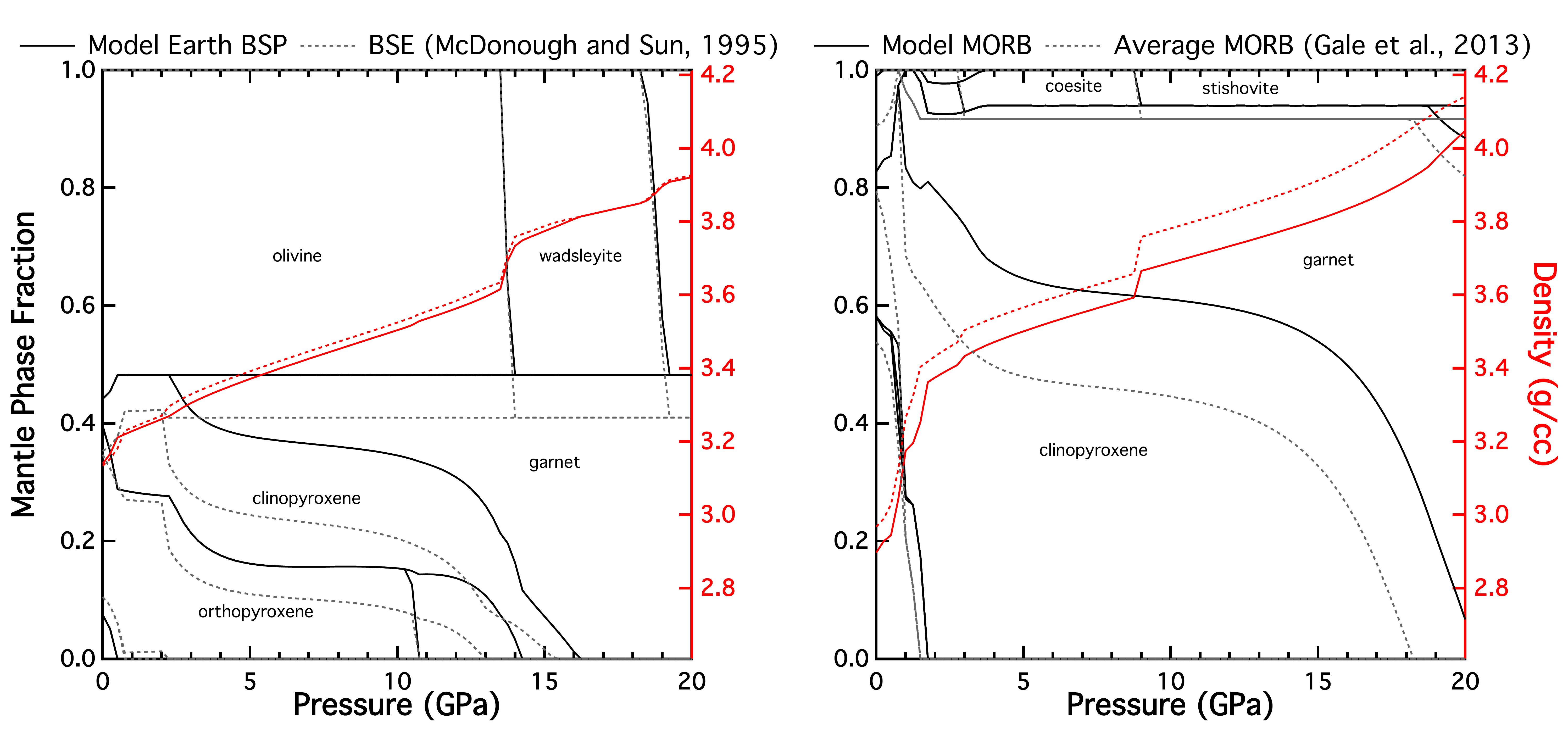}
    \caption{ Predicted phases (black) and densities (red) in this model (solid) compared to Earth reference compositions \citep[dashed,][]{McD03}for (a) the bulk silicate Earth and (b) melt-extracted basalt as a function of pressure. The model mantle is calculated along a 1600 K adiabat while the basalt is calculated along a 1200 K adiabat. While the simplified model presented here does not reproduce the Earth in detail, it does reproduce the correct minerals and pressures of phase transitions and magnitude of density discontinuities. Our model mantle under predicts the olivine fraction in the mantle (55\% vs. 60\%) while our model basalt predicts clinopyroxene at the expense of garnet relative to average basalt compositions under predicting the density by no more than 3\%. Garnet is the dense phase in the basalt-eclogite transition; this model therefore represents a lower bound on the likelihood of subduction. }
    \label{fig:phases}
\end{figure}
The condensation temperatures of the moderately volatile elements are highly correlated with their abundances in ordinary and CI chondrites \citep{Wai77}, indicating a condensation process that is both composition and thermally-driven. However, because of the lower condensation temperature of these elements relative to the refractories \citep[e.g.][]{Lodd03}, there is likely radial mixing of material within the disk. Recent planetary formation models \citep[e.g.][]{Raym06} show water ice mixing from beyond 2.5 AU into the inner Solar System, while the degree of mixing of the moderately volatile elements increases with the timescale of planet formation \citep{Bond10a,Bond10b}. Other disk models addressing the mixing of moderately volatile elements in the disk systematically over-predict the Na abundance in resulting planets compared to when attempting to reproduce the Earth's composition \citep{Matsu16}, likely due to neglecting not accounting for volatile loss due to melting during planetary formation.

\begin{deluxetable}{l|cc|cc|cc|cc}
\tablecolumns{9}
\tablecaption{Relative abundances of the most abundance elements considered in this model normalized to 100 atoms; Calculation of the BSP composition as separated from the planet composition is at 0.4 GPa and fO$_{2}$ = IQ-1.4 \citep{Jone86}, while the calculation of the MORB composition is assumed to be at 5\% melt fraction at 1 GPa  \citep{Mall09} and fO$_{2}$=QFM+0.4 \citep{Pres02}.}
\tablehead{\colhead{}&\colhead{Solar}&\colhead{Solar}&\colhead{Whole Earth}&\colhead{Whole Earth}&\colhead{BSE}&\colhead{BSE}&\colhead{MORB}&\colhead{MORB}\\
\colhead{Element}&\colhead{Actual$^{1}$}&\colhead{Condensed$^{\dagger}$}&\colhead{Actual$^{2}$}&\colhead{Model}&\colhead{Actual$^{2}$}&\colhead{Model}&\colhead{Actual$^{3}$}&\colhead{Model}}
\startdata
Mg&6.1&16.6&16.9&17.4&19.9&19.7&3.8-4.6&4.3\\
Si&5.8&15.8&15.3&14.4&15.9&16.4&20.0-17.7&19.1\\
Fe&5.1&13.8&15.3&14.4&2.4&2.4&2.8-3.9&2.4 \\
Al&0.4&1.1&1.5&1.2&1.8&1.4&6.3-6.8&6.6 \\
Ca&0.4&1.0&1.1&1.0&1.3&1.2&4.4-4.6&3.5 \\
Na&0.3&0.7&0.2&0.2&0.2&0.2&1.6-2.5&3.5 \\
O&82.0&51.0&49.6&51.3&58.5&58.7&60.0-61.2&60.5\\
\enddata
\tablecomments{$^{\dagger}$Assuming 22.8\% of Oxygen enters refractory phases per \citet{Unte17}\\$^{1}$\citet{Aspl05}\\$^{2}$\citet{McD03}\\$^{3}$\citet{Gale13}}
\label{tab:comparison}
\end{deluxetable}

\subsubsection{Planetary Differentiation Model}

We model two stages of differentiation from the condensed planet composition: separation of the iron-rich core from the bulk silicate planet (BSP) and crustal rock forming as a result of partial melting (exocrust) from adiabatic decompression of the BSP. This simplified two-stage model broadly reproduces the major element composition of the Earth, core-extracted bulk silicate Earth, and mid-ocean ridge basalt (MORB) from the Solar composition (Supplementary Table \ref{tab:comparison}). Differences between our simplified Earth and the true Earth lead to comparable differences in modal abundance of minerals in both MORB and Bulk Silicate Earth (BSE, Supplementary Figure \ref{fig:phases}). The resulting absolute difference in calculated densities vary by 0.6\% for the bulk silicate Earth and 2.6\% for the modeled MORB composition, with the depth-to-metamorphic transitions and the relative difference between the transitions for each composition indistinguishable. 

We model planetary differentiation through self-consistent thermodynamic modeling \citep{Ghio95,Ghio02} of cooling the bulk composition from a molten planet to calculate mineral and melt equilibria. The pressure and gas fugacity of the equilibrium calculations are chosen to best reproduce the fraction of differentiated Fe alloy and the composition of the BSE and MORB composition from the Sun's composition. We assume Si is primary the light element in the core \citep{Fisch15} to be consistent with the Mg/Si ratio of the Earth's mantle \citep{McD95,McD03,Unte16}. The melting of the mantle is calculated by assuming an adiabatic rise of material to lower pressures until the solidus temperature of the rock is reached. We calculate the primary crust composition to be that of the BSE at 5\% melt fraction, which is comparable to average mid-ocean ridge basalt (Supplementary Table \ref{tab:comparison}) after imposing a systematic correction to MgO to account for an over stabilization of clinopyroxene in MELTS \citep{Ghio02}. If no Si is present in the core, the mantle Si abundance would increase accordingly. This would enrich the melt in silica, forming a crust of buoyant andesite rather than basalt. We solve for the thermodynamically stable mineralogical host of each BSP and exocrustal composition as a function of pressure and temperature using the HeFESTo \citep{Stix05,Stix11}, open-source BurnMan \citep{Cott14,Unte16} and the Exoplanet-Pocketknife (available at https://github.com/ScottHull/Exoplanet-Pocketknife) software packages. Temperatures are assumed to increase adiabatically, calculated self-consistently with depth. The mantle adiabat is fixed at 1600 K, approximately the potential temperature of the Earth's mantle, while the average exocrust temperature is varied between 1000 K geotherms ($\Delta \rm{T} =600 K$) and 1600 K (thermally equilibrated, $\Delta \rm{T}=0$) From the mineralogy, compositions of the minerals, and modal abundance of each, density profiles of profile are calculated from the mineral-specific equations of state.

\begin{figure}
    \centering
    \includegraphics[width=\linewidth]{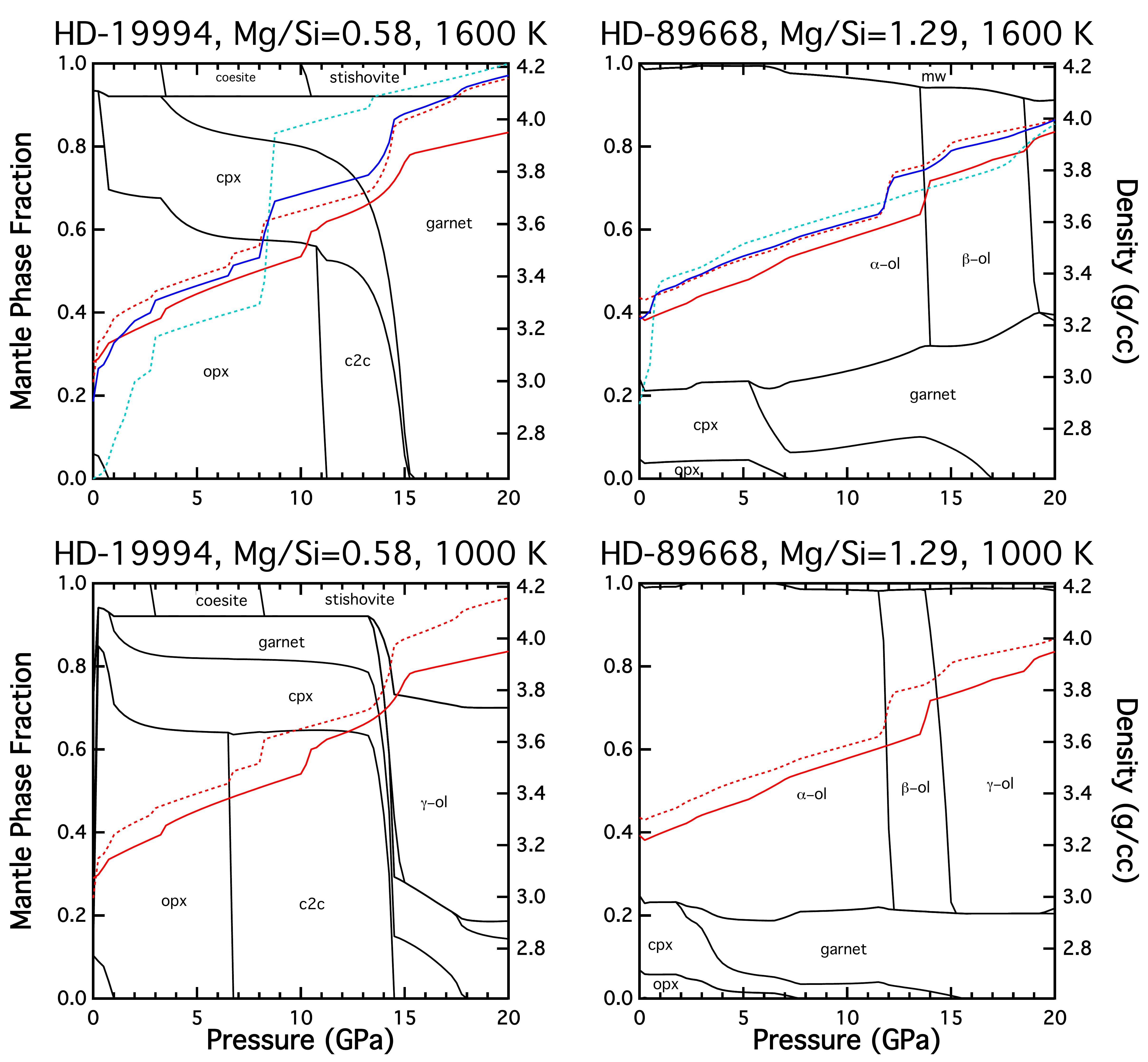}
    \caption{Predicted mantle mineralogies along 1600 K (top) and 1000 K (bottom) geotherms of hypothetical terrestrial planets about stars HD-19994 (left) and HD-89668 (right), representing extremes in Mg/Si. cpx = clinopyroxene, opx = orthopyroxene, c2c = c2/c structured-phase of opx, mw= magnesiowustite, $\alpha$-, $\beta$-, $\gamma$-ol = olivine, wadsleyite, and ringwoodite, respectively. Densities (right axis) of the mantle along a 1600 K geotherm (solid red) are compared to the densities of the inferred slab along a 1000 K geotherm (solid blue) are a function of the density of the exocrust (dashed cyan) and cold mantle lithosphere (dashed red). 
}
    \label{fig:nonEarth}
\end{figure}

\begin{figure}
    \centering
    \includegraphics[width=\linewidth]{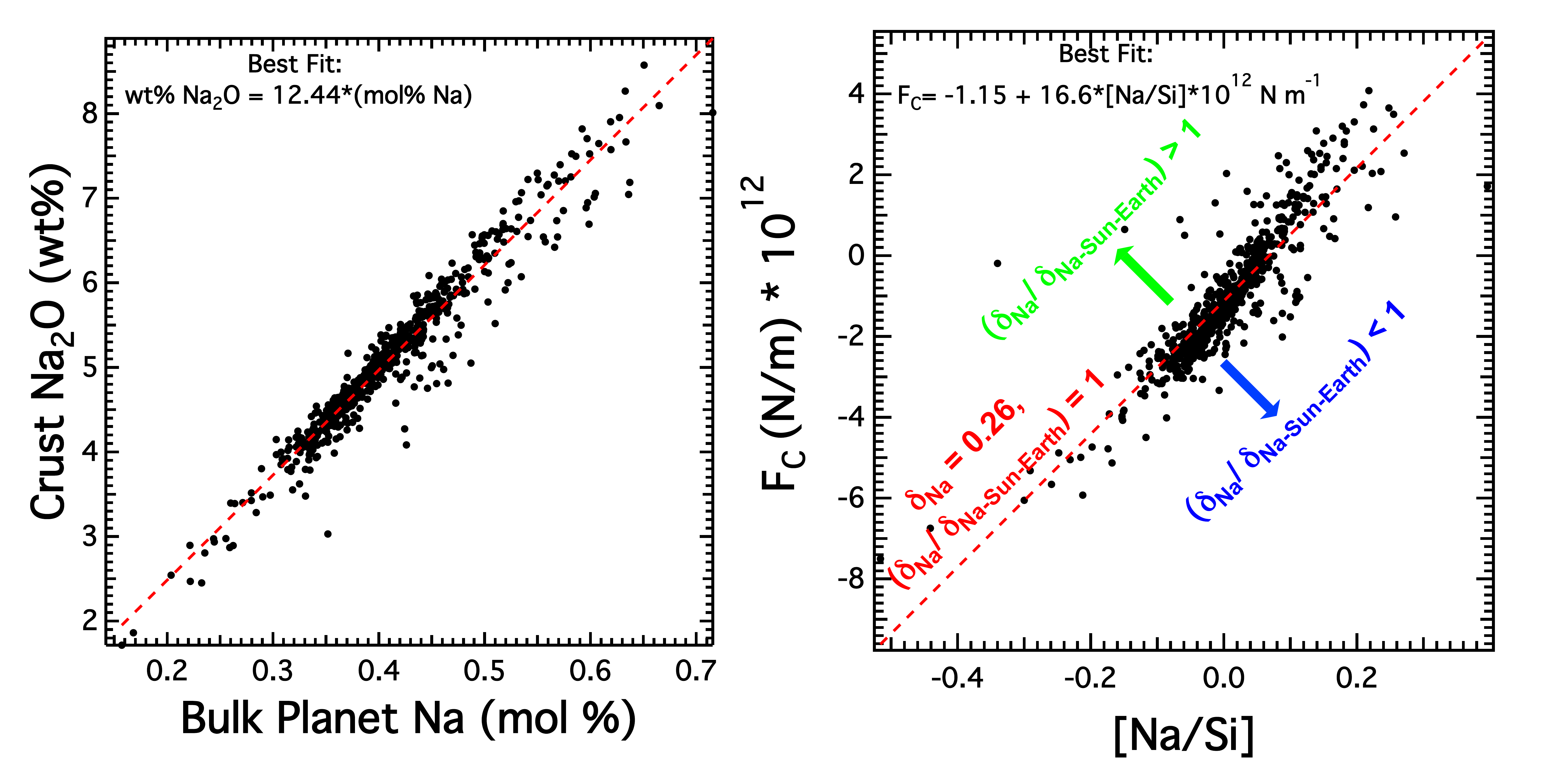}
    \caption{Crustal Na$_{2}$O wt\% composition as a function of molar fraction Na and chemical buoyancy force ($F_{c}$) as a function of stellar [Na/Si] for stars within the MELTS database. We adopt the Solar model of \citet{Aspl05} for calculation of [Na/Si]. $F_{c}$ is modeled assuming Solar Na is reduced by 74\% to the Earth abundance relative to Si \citep{Aspl05,McD03}. }
    \label{fig:fit}
\end{figure}

An additional consideration is the extent of melting for the formation of the exocrust, which is a function of both mantle composition and temperature. We find that 95\% of the sample set is at 5\% melt over a 65 K range, such that forming an exocrust of thickness comparable to that of the Earth occurs over a relatively narrow range. For lower mantle temperatures, little-to-no melt may be produced, while greater mantle temperatures will produce greater crustal thickness. For simplicity, we have considered a self-consistent adiabat applicable to the Earth with an Earth-like potential temperature (1600 K). Increasing the potential temperature increases the extent of melting, and consequently the thickness of melt-extracted crust. A greater extent of melting decreases Na$_2$O fraction in the melt for an approximately constant SiO$_2$ fraction when melt fraction $>$ 5\% \citep{Ghio02}, both effects increase the downward force due to chemical buoyancy for a given composition planet. The Earth's mantle contains trace quantities of both H$_2$O and CO$_2$, which affect the melting behavior of the passive upwelling mantle. Carbon increases the depth of initial melting in the decompression melting of mantle peridotite. This melting is at significantly greater depths than assumed in the models here, but contributes very small fractions of melt \citep[$<$0.3\%;][]{Dasg06} and therefore unlikely to significantly affect the major element chemistry assumed here. Similarly, water decreases the solidus temperature, increasing the depth of initial melting, but not significantly altering the total melt fraction \citep{Asim03} and decreases the total iron in the melt.

With a simple two-stage model, we do not address the chemical secular evolution of the mantle due to the extraction of continental crust. As such, we overestimate the Na$_{2}$O in the basaltic crust due to neglecting the abundance of Na$_{2}$O sequestered in the continental crust. For a continental crust with 3.6 wt\% Na$_{2}$O \citep{Rudn03}, reducing the BSP Na$_{2}$O abundance by about 5\%, reducing the Na$_{2}$O content of the resulting melt-extracted crust about 77\% of our predicted Na$_{2}$O content, consistent with typical MORB compositions. This suggests that early plate tectonic processes on Earth required initial secular variation in bulk composition of incompatible elements before modern plate tectonics could commence. 

The downward sinking force is quantified by the buoyancy of a potential sinking plate, made up of 10 km-thick exocrust and 50 km-thick lithospheric mantle. For simplicity, we adopt the BSP composition as the lithospheric mantle composition. The buoyancy force per unit length subducting plate, $F_{b}$, is
\begin{equation}
F_{b} = t \int_{0}^{d}\Delta\rho\left(\Delta x,h,\Delta T\right)g\left(h\right)dh
\end{equation}
where $\Delta\rho(\Delta x,h, \Delta T)$ is the density difference between the down-going plate and mantle compositions as a function of differences in the bulk compositions, $x$, depth, $h$, and the potential temperature difference between the mantle and slab, $\Delta T$, $g$ is the acceleration due to gravity, $d$ is the depth to the base of the Earth's transition zone (where the integrated density differences are a maximum), and $t$ is the slab thickness. While $g$ is a function of the planet mass, the depth to metamorphic transitions are pressure dependent, such that for a planet of equal core to mantle proportions, the product of surface gravity and the depth at 25 GPa varies by just 1.3\% over planets between 1 and 4.5 Earth masses. Therefore, we adopt Earth's surface acceleration due to gravity and Earth's pressure-depth relationship as a representative model valid for the calculation of net negative buoyancy in Super-Earths.

The magnitude of the chemical buoyancy force is closely tied to the abundance of Na$_{2}$O in the BSP due to its incompatibility in melting processes, 
\begin{equation}
    F_{c} \approx (-9.38+1.64*X_{Na_{2}O})*10^{12} \rm{N m^{-1}}
\end{equation}
where $X_{Na_{2}0}$ is the weight percent Na$_{2}$O in the planet's mantle ($X_{Na_{2}0}$ = 4.97 wt\% for our model Earth. 

\end{document}